\documentclass{ws-procs9x6mod}

\newcommand{\Nf}{N_{\!f}} 
 
\newcommand{\NA}{N_{\!A}} 
\newcommand{\MSbar}{\overline{\mbox{MS}}} 

\begin{document}

{\hspace{9.2cm}
{\bf LTH 777}} 

\title{RECENT RESULTS FOR YANG-MILLS THEORY RESTRICTED TO THE GRIBOV 
REGION\footnote{Talk presented at 9th International Conference on Path
Integrals - New Trends and Perspectives, Max Planck Institute for the Physics
of Complex Systems, Dresden, Germany, September 23-28, 2007}}

\author{J. A. GRACEY$^*$}

\address{Theoretical Physics Division, Department of Mathematical Sciences, 
University of Liverpool,
P.O. Box 147, Liverpool, L69 3BX, United Kingdom\\
$^*$E-mail: gracey@liv.ac.uk}

\begin{abstract}
We summarize recent results for the Gribov-Zwanziger Lagrangian which includes 
the effect of restricting the path integral to the first Gribov region. These 
include the two loop $\MSbar$ and one loop MOM gap equations for the Gribov 
mass.
\end{abstract}


\bodymatter

\section{Introduction}
The generalization of quantum electrodynamics to include non-abelian gauge
fields produces the asymptotically free gauge theory called quantum
chromodynamics (QCD) which describes the strong interactions. The natural forum
to construct the properly gauge fixed (renormalizable) Lagrangian with which to
perform calculations, is provided by the path integral machinery. For instance 
in the Landau gauge, which we concentrate on here, the Faddeev-Popov ghosts
naturally emerge as a consequence of the non-gauge invariance of the path
integral measure. Whilst the resulting Lagrangian more than adequately 
describes the ultraviolet structure of asymptotically free quarks and gluons 
the infrared behaviour has not been fully established. For instance, it is 
evident that as a result of confinement gluons and quarks cannot have 
propagators of a fundamental type. Over the last few years there has been 
intense activity into measuring gluon and ghost form factors using lattice 
methods and the Dyson Schwinger formalism. Denoting these respectively by 
$D_A(p^2)$ and $D_c(p^2)$ a general picture emerges in that there is gluon 
suppression with $D_A(0)$~$=$~$0$ and ghost enhancement where 
$D_c(p^2)$~$\sim$~$1/(p^2)^\lambda$ as $p^2$~$\rightarrow$~$0$ with 
$\lambda$~$>$~$0$. Such behaviour is not inconsistent with general 
considerations from confinement criteria\cite{1,2,3,4,5,6,7,8,9}. Ideally given
that these properties are now accepted, it is important that they can be 
explained from general field theory considerations. This was the approach of 
Zwanziger\cite{4,5,7,8} in treating the Gribov problem from the path integral 
point of view. Therefore we will briefly review the construction of the 
Gribov-Zwanziger Lagrangian before giving a summary of recent results of using 
it in the Landau gauge. 

\section{Gribov-Zwanziger Lagrangian}
Gribov pointed out\cite{1} that in non-abelian gauge theories it is not
possible to uniquely fix the gauge globally due to the existence of copies of
the gauge field. To handle this the path integral was restricted to the first
Gribov region, $\Omega$, where $\partial \Omega$ is defined by the place where
the Faddeev-Popov operator ${\cal M}$~$=$~$-$~$\partial^\mu D_\mu$ first
vanishes. Within $\Omega$ ${\cal M}$ is always positive and in the Landau 
gauge it is hermitian. Moreover $\Omega$ is convex and bounded\cite{3} and all 
gauge copies transit\cite{3} $\Omega$. Any copy in the subsequent regions 
defined by the other zeroes of ${\cal M}$ can be mapped into $\Omega$. Whilst 
the path integral is constrained to $\Omega$, within $\Omega$ there is a 
region, $\Lambda$, known as the fundamental modular region where there are no 
gauge copies and the gauge is properly fixed. Although $\Lambda$ is difficult 
to define, for practical purposes expectation values over $\Lambda$ or $\Omega$
give the same values\cite{10}. Consequently the gluon form factor is modified 
to $D_A(p^2)$~$=$~$(p^2)^2/[(p^2)^2+C_A\gamma^4]$ where $\gamma$ is the Gribov
mass, whence suppression emerges\cite{1}. The parameter $\gamma$ is not 
independent and satisfies a gap equation. The theory can only be interpreted as
a gauge theory when $\gamma$ takes the value defined in the gap equation. 
Thence computing the one loop ghost propagator, it is enhanced precisely when
the gap equation is satisfied\cite{1}. 

Gribov's revolutionary analysis was based on a semi-classical approach and then
Zwanziger\cite{4,5} extended it to a path integral construction by modifying 
the measure to restrict the integration region to $\Omega$ via the defining 
criterion known as the horizon condition, 
\begin{equation}
\int A^a_\mu(x) \frac{1}{\partial^\nu D_\nu} A^{a\,\mu}(x) ~=~ 
\frac{d N_A}{C_A g^2}
\label{hordef}
\end{equation}
where $d$ is the dimension of spacetime and $N_A$ is the adjoint representation 
dimension\cite{5}. For the Landau gauge the convexity and ellipsoidal 
properties of $\Omega$ allow one to modify the usual Yang-Mills Lagrangian to 
include the horizon condition, (\ref{hordef}), producing the non-local 
Yang-Mills Lagrangian\cite{4,5} 
\begin{equation}  
L^\gamma ~=~ -~ \frac{1}{4} G_{\mu\nu}^a 
G^{a \, \mu\nu} ~+~ \frac{C_A\gamma^4}{2} A^a_\mu \, \frac{1}{\partial^\nu 
D_\nu} A^{a \, \mu} ~-~ \frac{d \NA \gamma^4}{2g^2} ~. 
\label{nloclag}
\end{equation} 
Again (\ref{nloclag}) only has meaning when $\gamma$ satisfies (\ref{hordef})
which is equivalent to the Gribov gap equation. Finally the non-locality can
be handled by using localizing fields to produce the Gribov-Zwanziger
Lagrangian\cite{5} 
\begin{eqnarray} 
L^Z &=& L^{QCD} ~+~ \bar{\phi}^{ab \, \mu} \partial^\nu
\left( D_\nu \phi_\mu \right)^{ab} ~-~ \bar{\omega}^{ab \, \mu} \partial^\nu 
\left( D_\nu \omega_\mu \right)^{ab} \nonumber \\  
&& -~ g f^{abc} \partial^\nu \bar{\omega}^{ae}_\mu \left( D_\nu c \right)^b
\phi^{ec \, \mu} \nonumber \\
&& +~ \frac{\gamma^2}{\sqrt{2}} \left( f^{abc} A^{a \, \mu} \phi^{bc}_\mu ~+~ 
f^{abc} A^{a \, \mu} \bar{\phi}^{bc}_\mu \right) ~-~ 
\frac{d \NA \gamma^4}{2g^2} 
\label{gzlag}
\end{eqnarray} 
where $\phi^{ab}_\mu$ and $\omega^{ab}_\mu$ are localizing ghost fields with 
the latter anti-commuting. This Lagrangian is renormalizable\cite{7,11,12} 
and reproduces Gribov's one loop gap equation and ghost enhancement\cite{8}. 
For (\ref{gzlag}) the horizon condition equates to  
\begin{equation}
f^{abc} \langle A^{a \, \mu}(x) \phi^{bc}_\mu(x) \rangle ~=~ 
\frac{d \NA \gamma^2}{\sqrt{2}g^2} ~. 
\label{hordefgz}
\end{equation}

\section{Calculations}
As the Zwanziger construction has produced a renormalizable Lagrangian with 
extra fields incorporating infrared features without upsetting ultraviolet 
properties, such as asymptotic freedom, it is possible to extend the earlier 
one loop analysis\cite{1,8}. For instance in $\MSbar$ the two loop gap equation
results from (\ref{hordefgz}) after computing $17$ vacuum bubble graphs, 
giving\cite{13}, 
\begin{eqnarray} 
1 &=& C_A \left[ \frac{5}{8} - \frac{3}{8} \ln \left( 
\frac{C_A\gamma^4}{\mu^4} \right) \right] a \nonumber \\ 
&& +~ \left[ C_A^2 \left( \frac{2017}{768} - \frac{11097}{2048} s_2
+ \frac{95}{256} \zeta(2)
- \frac{65}{48} \ln \left( \frac{C_A\gamma^4}{\mu^4} \right) \right. \right. 
\nonumber \\ 
&& \left. \left. ~~~~~~~~~~~~+~ \frac{35}{128} \left( \ln \left( 
\frac{C_A\gamma^4}{\mu^4} \right) \right)^2 + \frac{1137}{2560} \sqrt{5} 
\zeta(2) - \frac{205\pi^2}{512} \right) \right. \nonumber \\
&& \left. ~~~~~+~ C_A T_F \Nf \left( -~ \frac{25}{24} - \zeta(2)
+ \frac{7}{12} \ln \left( \frac{C_A\gamma^4}{\mu^4} \right) \right. \right.
\nonumber \\ 
&& \left. \left. ~~~~~~~~~~~~~~~~~~~~~~~-~ \frac{1}{8} \left( \ln \left( 
\frac{C_A\gamma^4}{\mu^4} \right) \right)^2 + \frac{\pi^2}{8} \right) \right] 
a^2 +~ O(a^3) 
\label{gap2}
\end{eqnarray} 
where $s_2$~$=$~$(2\sqrt{3}/9) \mbox{Cl}_2(2\pi/3)$ with $\mbox{Cl}_2(x)$ the
Clausen function, $\zeta(n)$ is the Riemann zeta function and 
$a$~$=$~$\alpha_S/(4\pi)$. To appreciate the non-perturbative nature of 
$\gamma$ one can formally solve for it with the ansatz 
\begin{equation}
\frac{C_A \gamma^4}{\mu^4} ~=~ c_0 [ 1 + c_1 C_A \alpha_S ] 
\exp \left[ - \frac{b_0}{C_A \alpha_S} \right]
\end{equation}
giving 
\begin{equation}
b_0 ~=~ \frac{32\pi\left[3C_A - \sqrt{79C_A^2-32C_A T_F \Nf}\right]}
{[35C_A-16T_F\Nf]} 
\end{equation}
\begin{equation}
c_0 = \exp \!\! \left[ \frac{1}{[105C_A - 48 T_F \Nf]} \! \! 
\left[ 260 C_A - 112 T_F \Nf 
- \frac{[255C_A - 96 T_F \Nf] C_A}{\sqrt{79 C_A^2 - 32 C_A T_F \Nf}} \right]
\! \right] 
\end{equation}
and
\begin{eqnarray} 
c_1 &=& \left[ 8940981420 \sqrt{5} C_A^4 \zeta(2)
             - 11330632512 \sqrt{5} C_A^3 \Nf T_F \zeta(2)
\right. \nonumber \\
&& \left. +~ 4778237952 \sqrt{5} C_A^2 \Nf^2 T_F^2 \zeta(2)
             - 670629888 \sqrt{5} C_A \Nf^3 T_F^3 \zeta(2)
\right. \nonumber \\
&& \left. -~ 8060251500 \pi^2 C_A^4 - 109078793775 s_2 C_A^4 
\right. \nonumber \\
&& \left. +~ 7470477000 C_A^4 \zeta(2) + 19529637400 C_A^4
\right. \nonumber \\
&& \left. +~ 12730881600 \pi^2 C_A^3 \Nf T_F + 138232221840 s_2 C_A^3 \Nf T_F
\right. \nonumber \\
&& \left. -~ 29598076800 C_A^3 \Nf T_F \zeta(2) - 32025280640 C_A^3 \Nf T_F
\right. \nonumber \\
&& \left. -~ 7496478720 \pi^2 C_A^2 \Nf^2 T_F^2
             - 58293872640 s_2 C_A^2 \Nf^2 T_F^2
\right. \nonumber \\
&& \left. +~ 29503733760 C_A^2 \Nf^2 T_F^2 \zeta(2)
             + 19655024640 C_A^2 \Nf^2 T_F^2
\right. \nonumber \\
&& \left. +~ 1949368320 \pi^2 C_A \Nf^3 T_F^3
             + 8181596160 s_2 C_A \Nf^3 T_F^3
\right. \nonumber \\
&& \left. -~ 11318722560 C_A \Nf^3 T_F^3 \zeta(2)
             - 5351014400 C_A \Nf^3 T_F^3
\right. \nonumber \\
&& \left. -~ 188743680 \pi^2 \Nf^4 T_F^4
             + 1509949440 \Nf^4 T_F^4 \zeta(2)
+ 545259520 \Nf^4 T_F^4 \right] \nonumber \\
&& \times \frac{1}{46080 \pi [ 79 C_A - 32 T_F \Nf ]^{5/2} 
[ 35 C_A - 16 T_F \Nf ] \sqrt{C_A}} ~. 
\end{eqnarray} 
So in principle one could now compute with a gluon propagator which includes
renormalon type singularities. Further, with (\ref{gap2}) there is two loop 
ghost enhancement with the Kugo-Ojima confinement criterion\cite{9} precisely 
fulfilled at this order consistent with Zwanziger's all orders proof\cite{7}.
Also at one loop it has been shown\cite{14} that $D_A(0)$~$=$~$0$. The final 
quantity of interest is the renormalization group invariant effective coupling 
constant $\alpha^{\mbox{eff}}_S (p^2)$~$=$~$\alpha_S(\mu) D_A(p^2) 
\left( D_c(p^2) \right)^2$ which is believed to freeze at zero momentum. From
the $\MSbar$ one loop form factors it was shown\cite{14} that 
$\alpha^{\mbox{eff}}_S (0)$~$=$~$\frac{50}{3\pi C_A}$.

Whilst the previous expressions have all been in the $\MSbar$ scheme it is
worth considering other renormalization schemes such as MOM. Given that one
loop calculations\cite{14} produced exact form factors the derivation of the 
one loop MOM gap equation is straightforward, giving 
\begin{eqnarray}
1 &=& \left[ \frac{5}{8} + \frac{3}{8} \ln \left( 
\frac{C_A\gamma^4}{[C_A\gamma^4+\mu^4]} \right) - \frac{C_A\gamma^4}{8\mu^4} 
\ln \left( \frac{C_A\gamma^4}{[C_A\gamma^4+\mu^4]} \right) -
\frac{3\pi\sqrt{C_A}\gamma^2}{8\mu^2} \right. \nonumber \\ 
&& \left. + \left[ \frac{3\sqrt{C_A}\gamma^2}{4\mu^2} 
- \frac{\mu^2}{4\sqrt{C_A}\gamma^2} \right] \tan^{-1} \left[ 
\frac{\sqrt{C_A}\gamma^2}{\mu^2} \right] \right] C_A a + O(a^2) ~.
\label{momgap}
\end{eqnarray}
For later we formally define this as
$1$~$=$~$\mbox{gap}(\gamma,\mu,\mbox{MOM}) C_A a$~$+$~$O(a^2)$. Central to 
deriving this was the preservation of the Slavnov-Taylor identities in MOM.
For instance defining $Z_A$ and $Z_c$ from the respective gluon and ghost
$2$-point functions in MOM, then the coupling constant and $\gamma$
renormalization constants are already fixed and these must be used in computing
the horizon function. Given (\ref{momgap}) we have reproduced the one loop
ghost enhancement in MOM and the {\em same} freezing value for
$\alpha^{\mbox{eff}}_S (0)$. Since the numerical structure is different from 
the $\MSbar$ calculation we record the analogous\cite{14} computation is 
\begin{equation}
\alpha^{\mbox{eff}}_S (0) ~=~ \lim_{p^2 \rightarrow 0} \left[  
\frac{ \alpha_S(\mu) \left[ 1 - C_A \left( \mbox{gap}(\gamma,\mu,\mbox{MOM})
+ \frac{5}{8} - \frac{265}{384} \right) a 
\right] (p^2)^2 } 
{ C_A \gamma^4 \left[ 1 - C_A \left( \mbox{gap}(\gamma,\mu,\mbox{MOM}) 
- \frac{\pi p^2}{8 \sqrt{C_A} \gamma^2} \right) a \right]^2 } \right] 
\end{equation}
whence $\alpha^{\mbox{eff}}_S (0)$~$=$~$\frac{50}{3\pi C_A}$. 

\section{Discussion} 
To conclude we note that we have reviewed the path integral construction of
Zwanziger's localised renormalizable Lagrangian for the Landau gauge which 
incorporates the restriction of gauge configurations to the first Gribov 
region. A picture emerges of the infrared structure which is consistent with
the gluon being confined. Crucial to the analysis was the geometry of the
Gribov region. This can be appreciated from another point of view given recent
work in trying to extend the path integral construction to other 
gauges\cite{15,16,17}. For linear covariant gauges other than Landau the 
Fadeev-Popov operator is not hermitian\cite{15} and convexity of the Gribov
region is only valid when the covariant gauge fixing parameter is 
small\cite{15}. Moreover, given that the Faddeev-Popov operator in this 
instance would involve the transverse part of the gauge field then the 
non-local operator of (\ref{nloclag}) would itself contain a non-locality in 
the covariant derivative\cite{15}. Another example is the construction of a 
Gribov-Zwanziger type Lagrangian for $SU(2)$ Yang-Mills fixed in the maximal 
abelian gauge\cite{16,17}. Whilst a localised renormalizable Lagrangian 
analogous to (\ref{gzlag}) can be constructed the algebraic renormalization 
analysis demonstrates that there is an additional free parameter which has no 
analogue in the Landau gauge\cite{17}. Given these recent considerations it 
would seem therefore that in the Gribov context the Landau gauge is peculiarly 
special. 

\section*{Acknowledgements}
The author thanks Prof. S. Sorella, Prof. D. Zwanziger and Dr D. Dudal for 
useful discussions concerning the Gribov problem.

\end{document}